\documentclass[prl,twocolumn,showpacs,amsmath,amssymb]{revtex4-2}

\usepackage{graphicx}
\usepackage{dcolumn}
\usepackage{bm}

\begin{document}

\title{Ultimate plasma wakefield acceleration with 400 GeV proton driver}

\author{Konstantin Lotov}
\email{K.V.Lotov@inp.nsk.su}
\author{Petr Tuev}%
\affiliation{%
Budker Institute of Nuclear Physics SB RAS, Novosibirsk, Russia\\
Novosibirsk State University, Novosibirsk, Russia
}%

\date{\today}

\begin{abstract}
A new regime of proton-driven plasma wakefield acceleration is discovered, in which the plasma nonlinearity increases the phase velocity of the excited wave compared to that of the protons. If the beam charge is much larger than minimally necessary to excite a nonlinear wave, there is sufficient freedom in choosing the longitudinal plasma density profile to make the wave speed close to the speed of light. This allows electrons or positrons to be accelerated to about 200\,GeV with a 400\,GeV proton driver.
\end{abstract}


\maketitle

Plasma wakefield acceleration driven by proton beams combines the advantages of the high accelerating rates achievable in plasmas and the single-stage acceleration possible due to the high energy of available proton beams \cite{NatPhys5-363,PoP18-103101,RAST9-85}. However, because of the large mass of protons, the energy of accelerated particles (electrons, positrons or muons) can approach the initial proton energy only if the latter is much higher than 1 TeV \cite{PRST-AB13-041301}. The problem comes from dephasing of accelerated particles and the plasma wave. The phase velocity of the wave is close to the driver velocity and, for sub-TeV protons, is substantially lower than the velocity of accelerated particles, which quickly approaches the speed of light $c$. However, multi-TeV proton beams currently exist only at CERN and will not be available for plasma acceleration experiments in the near future. Therefore, it is interesting to find out to what energy electrons can be accelerated using sub-TeV proton beams, which are more accessible.

\begin{table}[tb]
\caption{Parameters of the plasma and beams.}\label{t1}
\begin{ruledtabular}
\begin{tabular}{ll}
  Parameter and notation & Value \\ \hline
  Initial plasma density, $n_0$ & $7 \times 10^{14}\,\text{cm}^{-3}$ \\
  Plasma skin depth, $k_p^{-1}$ & 0.2\,mm \\
  Plasma radius & 1.4\,mm \\
  Plasma ion mass number & 85 \\
  \textit{Proton beam:} &\\
  Population (charge) & $3\times 10^{11}$ (48\,nC) \\
  Root-mean-square (RMS) length & 7\,cm \\
  RMS radius & 0.2\,mm \\
  Energy & 400\,GeV \\
  RMS energy spread & 0.035\,\% \\
  Normalized emittance & 2.2\,mm\,mrad \\
  \textit{Seed electron beam:} &\\
  Population (charge) & $3.125\times 10^9$ (500\,pC)\\
  RMS length, $\sigma_{ze}$ & 0.66\,mm \\
  RMS radius, $\sigma_{re}$ & 0.25\,mm \\
  Energy, $W_e$ & 18\,MeV \\
  Normalized emittance & 4\,mm\,mrad \\
  \textit{Witness electron / positron beam:} &\\
  Population (charge) & $0.625\times 10^9$ (100\,pC)\\
  RMS length & 30\,$\mu$m / 45\,$\mu$m  \\
  RMS radius & 5.75\,$\mu$m \\
  Energy & 150\,MeV \\
  Normalized emittance & 2\,mm\,mrad
\end{tabular}
\end{ruledtabular}
\end{table}

As a baseline case, we consider the 400\,GeV proton beam of Super Proton Synchrotron (SPS), which is already used for plasma acceleration in the Advanced WAKefield Experiment (AWAKE) (Table~\ref{t1}) \cite{Nat.561-363,PPCF60-014046,NIMA-829-76}. In earlier simulations of the SPS driver, the maximum energy of  accelerated electrons was about 10\,GeV \cite{PoP18-103101} or 55\,GeV \cite{PBC}. Here we show that, using plasma nonlinearity and special longitudinal plasma density profiles, it is possible to accelerate electrons or positrons beyond 200\,GeV.

Proton beams from modern synchrotrons can excite a plasma wave only after conversion into a train of short micro-bunches either before \cite{RuPAC16-303} or inside the plasma, as a result of the seeded self-modulation (SSM) process \cite{PRL104-255003,PoP22-103110,PRL122-054801,PRL122-054802}. If the self-modulation is seeded by a co-propagating short laser pulse, as in the first run of AWAKE \cite{Nat.561-363}, then only approximately one eighth of the beam charge remains in the micro-bunches and resonantly drives the wave: the leading half of the beam moves in a neutral (not yet ionized) gas ahead of the laser pulse, and only a quarter of the trailing half survives, which falls into both focusing and decelerating phase of the wave. Nevertheless, this charge is sufficient to drive the wave to the limit determined by the nonlinear elongation of the wave period \cite{PoP21-083107}. The bunches go out of resonance with the wave, and the wave stops growing when the wakefield amplitude reaches 40--50\% of the wavebreaking field $E_0 = m c^2 k_p/e$ \cite{PoP20-083119}, where $m$ is the electron mass and $e$ is the elementary charge.

In the second run of AWAKE, the self-modulation will be seeded with a short electron beam \cite{JPCS1596-012008}. This method is twice more charge-efficient, since the entire beam is micro-bunched. In addition, the longitudinal profile of the plasma density will have a small step in the self-modulation region, so that the micro-bunches will be in radial equilibrium with the excited wave and, therefore, can propagate over a long distance without being destroyed by transverse forces \cite{PoP18-024501,PoP22-103110}.
The redundancy in beam charge gives freedom to optimize the bunch train by varying plasma density profiles at the self-modulation stage. Bunch trains produced at different profiles may initially drive a wave of the same large amplitude, but behave differently at long propagation distances, and the best ones are very effective.

\begin{figure}
\includegraphics{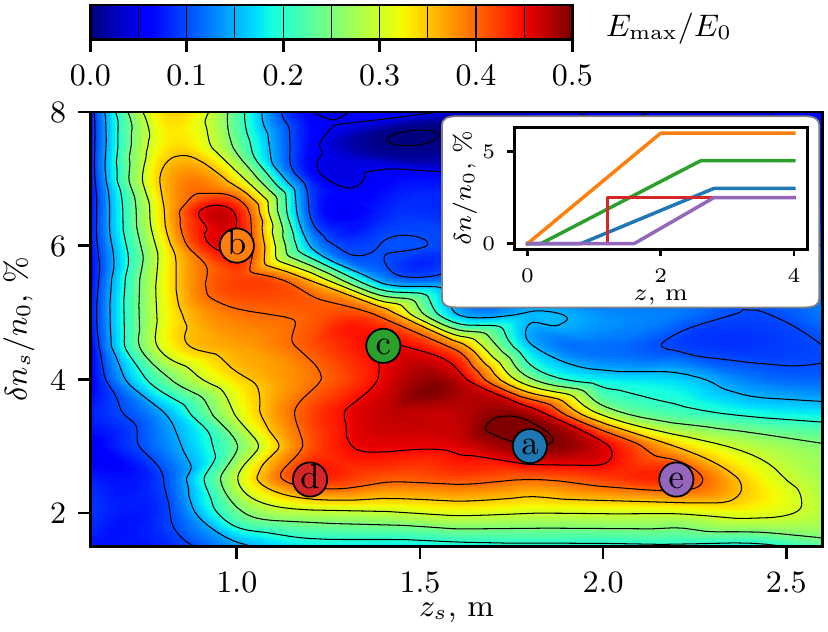}
\caption{The maximum longitudinal wakefield $E_\text{max}$ attainable after 20 meters of beam propagation in the plasma for different locations $z_s$ and magnitudes $\delta n_s$ of the density step. The step width is optimized for the strongest field. The letters in colored circles indicate the plasma density profile variants studied, and the correspondingly colored lines in the inset represent the profile of each variant.}
\label{fig1-map}
\end{figure}
\begin{figure}
\includegraphics{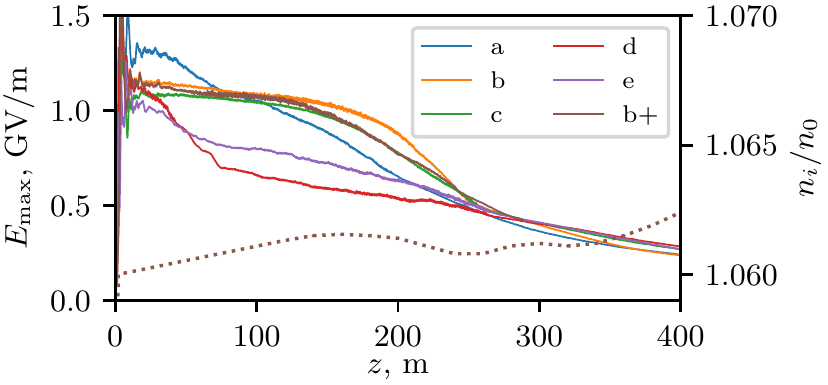}
\caption{The maximum longitudinal wakefield $E_\text{max}$ versus the propagation distance $z$ for the selected variants. The dotted line shows the plasma density profile $n_i(z)$ for the variant `b+', which coincides with the variant `b' at $z < 2$\,m and deviates from the constant value at $z > 2$\,m so that the phase velocity at $k_p \xi \approx -1300$ is equal to $c$.}
\label{fig2-gmaxf}
\end{figure}

We adhere to the same set of parameters as in Ref.~\cite{PPCF62-115025} (Table~\ref{t1}), for which we already know the relation between the density profiles and the wakefields excited after passing the first 20~meters (Fig.\,\ref{fig1-map}). The simulations are performed with the quasi-static 2d3v axisymmetric code LCODE \cite{PRST-AB6-061301,NIMA-829-350}. The grid step is $0.005 k_p^{-1}$ for both the radial coordinate $r$ and the longitudinal (co-moving) coordinate $\xi = z - ct$. The time step for the proton beam is $100 / (k_p c)$, the plasma state is calculated every $200 k_p^{-1}$. We select several typical points in Fig.\,\ref{fig1-map}, which correspond to different bunch trains, and study the long-term evolution of these trains. Larger steps produce bunch trains with a slower decaying wakefield (Fig.\,\ref{fig2-gmaxf}). Analysis of the wakefield growth along the beam shows that in these bunch trains, the micro-bunches at the beam leading edge work more efficiently and bring the amplitude to the limit faster. In turn, with smaller steps, the wakefield grows more slowly, and a larger number of micro-bunches contributes to this growth.

\begin{figure}
\includegraphics{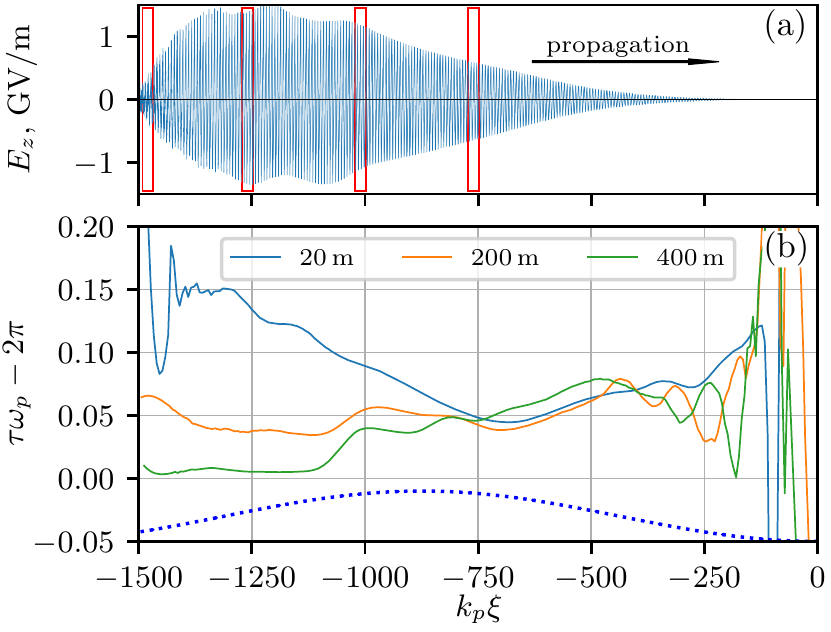}
\caption{(a) Growth of the longitudinal electric field $E_z$ along the proton beam at $z=20$\,m, variant `a'. The red rectangles indicate the areas detailed in Fig.\,\ref{fig4-phases}. (b) Variation of the wakefield period $\tau$ along the beam at different propagation distances, where $\omega_p = k_p c \sqrt{1 + \delta n / n_0}$ is the local plasma frequency. The dotted line shows the original beam shape.}
\label{fig3-period}
\end{figure}
\begin{figure}
\includegraphics{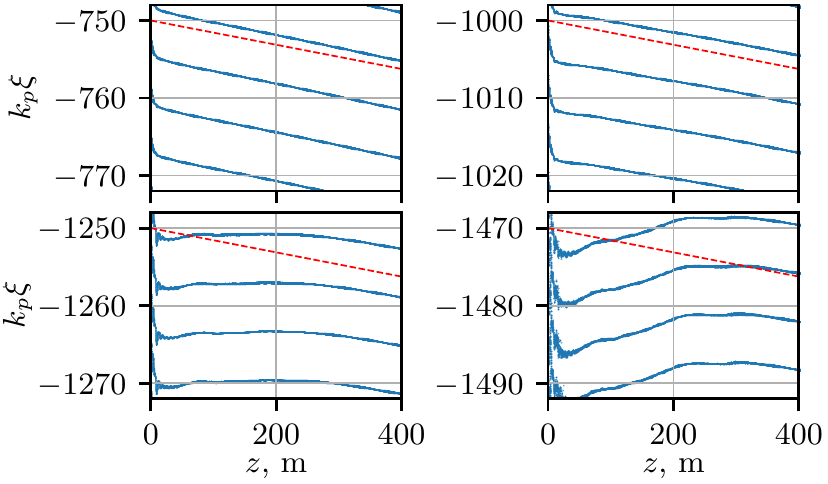}
\caption{The location of odd field zeros in the co-moving frame as a function of the propagation distance $z$ in the four intervals marked in Fig.\,\ref{fig3-period}(a). The slope of the dashed lines corresponds to the velocity of 400\,GeV protons.}
\label{fig4-phases}
\end{figure}

The initial nonlinearity of the wave leads to a specific behavior of the wakefield phase. While in most of the beam, the phase velocity of the wave is close to the velocity of 400\,GeV protons, there is a region at the back of the beam where it approaches the speed of light or even higher [Figs.\,\ref{fig3-period}(a) and \ref{fig4-phases}]. This is because the wave period is initially longer than the plasma period [Fig.\,\ref{fig3-period}(b)]. As the beam propagates in the plasma, the shape of micro-bunches slightly change, which leads to a decrease of the wave amplitude. In turn, this causes the wavelength to shorten, and the constant phase points move forward relative to the beam. At some distance from the beam head, this effect balances the slowness of the proton beam, and this location becomes suitable for long-term electron acceleration.

There is a competing effect that could also lead to similar behaviour of the wave phase. The period of forced oscillations in a growing wave is longer than the period of free oscillations, because each micro-bunch not only increases the wave amplitude, but also shifts the phase backward \cite{PoP22-103110,PPCF60-024002}. The wave period at the leading half of the beam is longer precisely because of this effect [Fig.\,\ref{fig3-period}(b)]. Hypothetically, as the micro-bunches deplete, the wave period could become shorter. However, a detailed examination \cite{PoP18-103101} shows that the micro-bunches do not deplete, but move into regions of a small longitudinal field, where they no longer increase the wave amplitude, but still affect the period. Thus, this effect cannot be responsible for the observed behavior of the wave phase.

\begin{figure}
\includegraphics{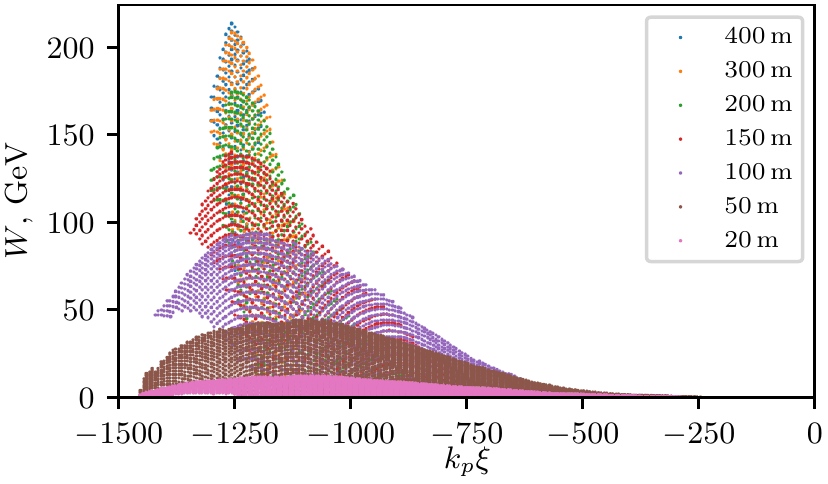}
\caption{The estimated electron energy $W(\xi)$ at different propagation distances for the variant `a', which provides the maximum energy gain in the uniform plasma.}
\label{fig5-prediction1}
\end{figure}
\begin{figure}
\includegraphics{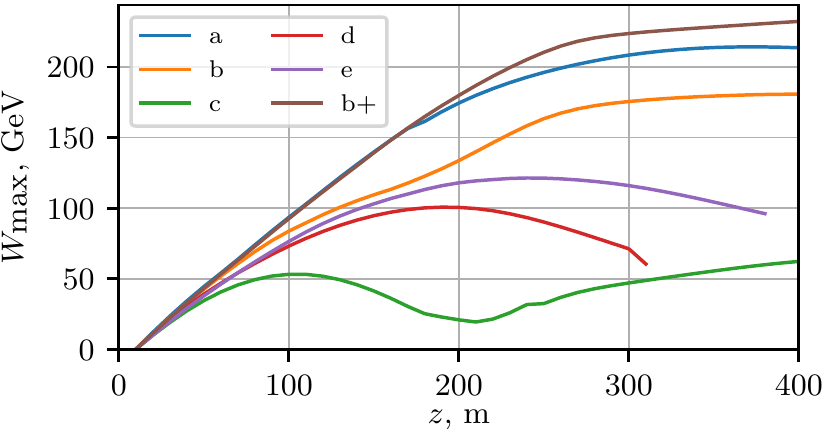}
\caption{The estimated maximum witness energy $W_\text{max} (z)$ for the considered variants.}
\label{fig6-prediction2}
\end{figure}

Finding the optimum location for witness electrons by directly simulating their acceleration over a long distance is computationally demanding, so we use a simplified approach. We find the coordinate and amplitude of local wakefield maxima, approximate the field between them by a cosine function, and calculate the witness energy gain, assuming that the electrons are always at $\xi = \text{const}$ and $r=0$. If the electron enters the defocusing phase of the wave or if the electron energy approaches the rest energy, the electron is excluded from consideration. The resulting energy gain gives a general idea of where acceleration is most effective (Fig.\,\ref{fig5-prediction1}) and which of the considered variants is the best (Fig.\,\ref{fig6-prediction2}). Although the field amplitude at $k_p \xi > -1000$ is also large [Fig.\,\ref{fig3-period}(a)], the energy gain over 100\,GeV is possible only in a narrow interval of $\xi$ where the  phase velocity is high (Fig.\,\ref{fig5-prediction1}). The variants with a slower decaying wakefield (`a' and `b' in Fig.\,\ref{fig2-gmaxf}) yield higher electron energies (Fig.\,\ref{fig6-prediction2}).

We also observe that at some potentially efficient regimes, the energy growth rate is not constant (variant `b' in Fig.\,\ref{fig6-prediction2}). The reason is that the wakefield phase is not perfectly stationary in the co-moving frame, and the fastest electrons enter the defocusing phase and disappear. This problem can be solved by slightly adjusting the plasma density along the beam path (dotted line in Fig.\,\ref{fig2-gmaxf}), resulting in even greater energy gains (variant `b+' in Fig.\,\ref{fig6-prediction2}). Small density variations, however, affect the efficiency of wave drive by proton micro-bunches (Fig.\,\ref{fig2-gmaxf}), so the density adjustment can only correct small phase errors, but not the slowness of the driver.

The effect of density variations on phase behavior becomes stronger as the distance between the witness and the driver increases. In our case, however, this solution does not work, since the wakefield lifetime, depending on the regime, is limited either by the ion motion \cite{PRL109-145005,PoP21-056705} or by the appearance of halo electrons \cite{PPCF63-055002} and does not exceed the driver duration [Fig.\,\ref{fig3-period}(a)].

\begin{figure}
\includegraphics{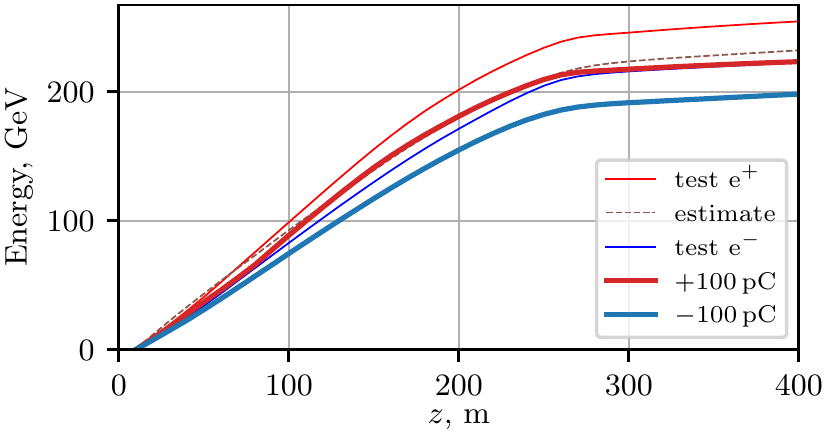}
\caption{The maximum energy gain of test electrons and positrons (thin lines), the average energy gain of 100\,pC electron and positron bunches (bold lines) and the energy gain estimate $W_\text{max}$ (dashed line) for the variant `b+'.}
\label{fig7-egain}
\end{figure}
\begin{figure}
\includegraphics{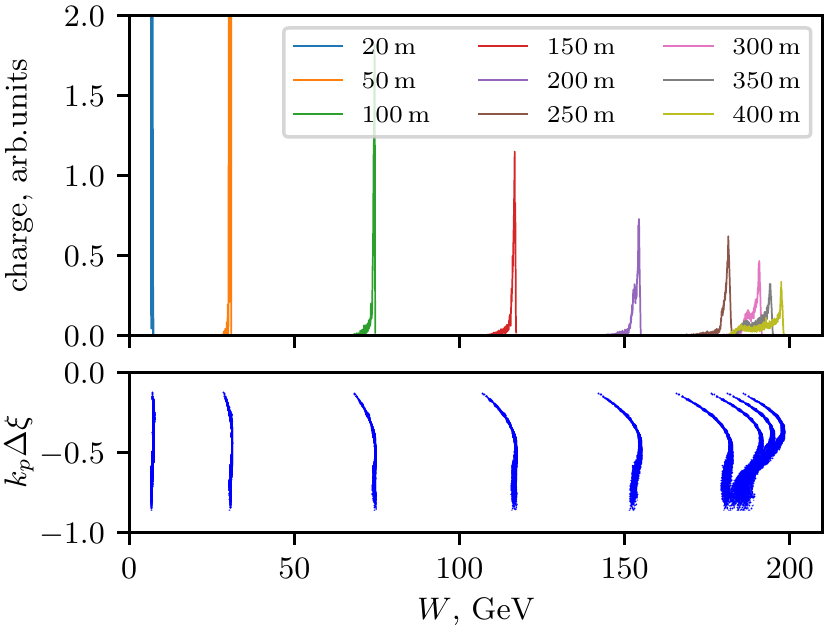}
\caption{(top) The energy spectra of the 100\,pC electron bunch at different propagation distances; (bottom) the corresponding longitudinal phase portraits of the bunch.}
\label{fig8-espectra}
\end{figure}

To confirm the possibility of high energy gains, we simulate the acceleration of electron and positron beams in the excited wakefields (Fig.\,\ref{fig7-egain}). The beams are injected (added to the simulation domain) after the proton beam has passed 10 meters and self-modulated, and the field profile in the co-moving frame stabilized. The length and precise position of the witness beams relative to the wave are optimized to reduce the final energy spread while maintaining a high acceleration rate and conserving the injected charge (Table~\ref{t1}). When simulating witness dynamics, we use an energy-dependent substepping \cite{PPCF62-115025}.

The maximum energy gain of the test (low charge) beams is close to the estimate, which justifies our simplified approach. For the higher charges, the energy gain is smaller, but still about 200\,GeV. The acceleration is almost charge-symmetric. The lowest energy spread achieved for these Gaussian beams is about 1\% for electrons and 5\% for positrons. This is possible due to beam loading, which flattens the average accelerating field (Fig.\,\ref{fig8-espectra}).

The effective acceleration length in our case is about 250\,m. It can be reduced by operating at higher plasma density, since the accelerating field scales as $\sqrt{n_0}$.

We did not analyze the emittance of the accelerated beams, since this study requires a much finer grid and shorter time steps \cite{PoP25-093112}. In our simulations, the emittance increases due to the numerical field noise, which leads to radial bunch expansion and an unphysical growth of the slice energy spread.

The optimum density profiles found in our simulations are by no means universal. For real beams in experiments, it is necessary to re-optimize the parameters of the density step. But this search is quite feasible, as there are only three parameters to configure (location, length, and magnitude of the density step). Fine-tuning the density profile, as was done for the variant `b+', may not be so easy and will probably require advanced optimization algorithms. However, this is not absolutely necessary, as it insignificantly increases the energy gain compared to the uniform plasma.

This study was supported by the Russian Foundation for Basic Research (RFBR) project 19-02-00243. Upgrade of LCODE for simulating long-term witness acceleration in plasmas of varying density was supported by RFBR project 19-31-90030. Simulations were performed on HPC-cluster "Akademik V.M. Matrosov" \cite{matrosov}.

\end{document}